\documentclass[11pt]{article}
\usepackage{epstopdf}
\usepackage{color}
\usepackage{subfigure}
\usepackage{amsmath}
\usepackage{amssymb}
\usepackage{graphicx,color}
\usepackage{cite}
\usepackage{enumerate}
\usepackage{amsthm}
\usepackage{amsfonts,mathrsfs}
\usepackage{geometry}
\usepackage{ifthen}
\usepackage{graphicx}
\usepackage{float}
\usepackage{bm}
\usepackage[utf8]{inputenc}
\usepackage{amsfonts}\usepackage[colorlinks,bookmarksopen,bookmarksnumbered,citecolor=blue, linkcolor=blue, urlcolor=blue]{hyperref}

\begin{document}

\title{\bf Two imaginarity monotones induced by unified $(\alpha,\beta)$-relative entropy}

\vskip0.1in
\author{\small Chuanfa Wu$^1$, Zhaoqi Wu$^1$\thanks{Corresponding author. E-mail: wuzhaoqi\_conquer@163.com}\\
{\small\it  1. Department of Mathematics, Nanchang University, Nanchang 330031, P R China}\\
}

\date{}
\maketitle

\noindent {\bf Abstract} {\small }\\
\indent Complex numbers play a pivotal role in both mathematics and physics, particularly in quantum mechanics, and are extensively utilized to depict the behavior of microscopic particles. Recognizing the significance of complex numbers, a framework of imaginarity resource theory has recently been established. In this work, we propose two types of imaginarity monotones induced by the unified $(\alpha,\beta)$-relative entropy and investigate their properties. Moreover, we give explicit examples to illustrate our results.\\


\noindent {\bf 1 Introduction} \\ \hspace*{\fill}\\
\indent The pivotal and indispensable role of complex numbers in quantum mechanics is undeniable, exerting profound and widespread influence. Complex numbers are important in building the framework of quantum mechanics. By using complex numbers, quantum states and wave functions can accurately describe the behavior of microscopic particles\cite{LJ,KR,GD}. In the dynamic evolution of quantum mechanics\cite{ASA}, complex numbers are particularly crucial. As an integral part of the Schrödinger equation \cite{RDS,SW,JG,BS,HK,RD,DJXM}, the introduction of the imaginary unit $\mathrm{i}$ not only accurately depict the evolutionary track of the wave function over time \cite{BBI,TFJ,SSHJM,ADCRS,HRHA,GSNZ,AANT}, but also endows essential phase information \cite{FDL,HZAV,SJDJ}. This phase information holds immeasurable value in comprehending unique quantum phenomena like interference \cite{WWRS,PXSJ,ZYZW} and diffraction \cite{ASKE,KSHH,QKWK}, while also laying a solid theoretical foundation for the advancement of cutting-edge technologies such as quantum computing \cite{BGCI,LUO5,VDWJ} and quantum communication \cite{HXMZ,CASC,CMCL}. In terms of experiments, the application of complex numbers in quantum mechanics has been thoroughly validated. By measuring the real and imaginary parts of the wave function \cite{BALK,TC,PLMR,WDJF,ASL,JMMJ,RMG,RRKA,WS}, the accuracy of quantum theory prediction is verified. This further reveals the mysteries of the quantum world. Therefore, we can gain a more comprehensive grasp of the operational laws of the microscopic realm by studying the imaginary numbers.\\
\indent In light of the pivotal role of complex numbers in quantum
mechanics, Hickey and Gour\cite{HAGG} proposed a landmark approach
in 2018, treating the imaginary component of quantum states as a
resource for investigation and subsequently constructing a
comprehensive framework, in which free states and free operations
has been elucidated. Many imaginarity quantifiers, such as the trace
norm of imaginarity \cite{HAGG}, the fidelity of imaginarity
\cite{WKKT0}, the weight of imaginarity \cite{XSGJ}, the relative
entropy of imaginarity \cite{XSGJ}, the robustness of imaginarity
\cite{WKDK}, the $l_1$ norm of imaginarity \cite{CQTG}, the
geometric-like imaginarity \cite{GMLB}, have been introduced and
studied. The relationship between the relative entropy of
imaginarity and the $l_1$ norm of imaginarity and the quantum state
order under different quantum channels have been revealed
\cite{CQTG}.

The above imaginarity measures of a quantum state $\rho$ are defined
by taking the minimum over all free states $\sigma$ for some
distance measures between states $\rho$ and $\sigma$. It is also
possible to define the measures by the distance from a quantum state
to its own conjugate, such as the Tsallis relative entropy of
imaginarity \cite{XJ}, the $\alpha$-$z$ R\'enyi relative entropy
imaginarity \cite{CXLQ} and the Tsallis relative operator entropy of
imaginarity \cite{CXLQ}. The problem of imaginarity of Gaussian
states \cite{XJ,XJ1} and Kirkwood-Dirac imaginarity have also been
considered\cite{FYGZ}.

In manipulating imaginarity resource, the main question is that can
any two given quantum states be converted into each other via free
operations. Sufficient and necessary conditions for the validity of
conversion under the circumstance of any two pure states\cite{HAGG}
and any two qubit states\cite{WKDK} have been obtained. Moreover,
imaginarity resource in distributed scenarios has been
discussed\cite{WKKT1}. Like coherence resource
\cite{BTRC,MDPT}, the freezing imaginarity of quantum state
\cite{HSZB} and the nonlocal advantages of imaginarity \cite{WZFS}
have been investigated. The connections between coherence and
imaginarity have also been explored\cite{ZLLN0,ZLLN1}.

The remainder of this paper is arranged as follows. In Section 2, we
review the fundamental concepts of the quantum imaginary resource
theory. In Section 3, we propose two imaginarity monotones induced
by the unified-$(\alpha,\beta)$ relative entropy and explore their
properties. In Section 4, we give some examples to illustrate our
results. The conclusion is provided in Section 5.

\vskip0.1in

\noindent {\bf   2 Preliminaries}\\\hspace*{\fill}\\
  \indent We denote by $\mathcal{H}$ a $d$-dimensional Hilbert space, $\{|j\rangle \}_{j=0}^{d-1}$ a fixed orthonormal basis in  $\mathcal H $, $\mathcal D( \mathcal H) $ the set of density operators (quantum states) $\rho$ acting on $  \mathcal H $ and  $\rho^*$ the conjugate of $\rho$. We recall the framework of the imaginarity resource theory.\\
  \indent  {\bf Definition 1} \cite{HAGG}\quad Let $ \rho $ be a quantum state on $  \mathcal H $. $ \rho $ is a free state or real state, if
\begin{equation}\label {eq1}
\rho=\sum_{jk} \rho_{jk} \ |j \rangle  \langle k|,
\end{equation}
where ${\rho_{jk}} \in  \mathbb{R} $ for arbitrary $j$,$k \in \{ 0,1,\dots,d-1\}$. We denote the set of all real states by $ \mathcal F $.\\
\indent It can be easily obtained that $ \rho \in \mathcal F $ if and only if $\rho $ is symmetric, that is, $  \rho^T=\rho $, where $\rho^T$ denotes the transpose of $\rho$.\\
\indent {\bf Definition 2}\cite{HAGG}\quad Let $ \Lambda $ be a quantum operation, $\rho $ be a quantum state and $ \Lambda (\rho)= \sum_{j} K_{j} \rho K_{j}^{\dagger} $. $ \Lambda $ is a free operation or real operation if
\begin{equation}\label {eq2}
\langle m|K_{j}| n \rangle \in \mathbb {R}
\end{equation} for arbitrary $j$ and $m,n \in \{0,1,\dots,d-1 \}$.\\
\indent {\bf Definition 3} \cite{HAGG}\quad An imaginarity measure is a functional $M$: $\mathcal D(\mathcal H) \to [0,\infty)$ satisfying the following properties:\\
(M1) Nonnegativity. $M(\rho) \ge 0$, $M(\rho)=0$ if and only if $\rho \in \mathcal F$.\\
(M2) Imaginarity monotonicity. $M(\Lambda (\rho)) \le M(\rho)$ whenever $\Lambda $ is a free operation.\\
(M3) Strong imaginarity monotonicity. $M(\rho) \ge  \sum_{j}p_{j}M(\rho_{j})$, where $p_{j}=\mathrm{tr}(K_{j} \rho K_{j}^{\dagger})$, $ \rho_{j}=K_{j} \rho K_{j}^{\dagger}/p_{j}  $ with free Kraus operators $K_{j}$.\\
(M4) Convexity. $\sum_{j}p_{j}M(\rho_{j}) \ge M(\sum_{j}p_{j} \rho_{j}) $ for any set of states $\{\rho_{j}\}$ and any  $p_{j} \ge 0$ with $\sum_{j}p_{j}=1 $.\\
\indent Besides, the following property (M5) has also been introduced.\\
(M5) Additivity. $M(p\rho_{1} \oplus (1-p)\rho_{2})=pM(\rho_{1})+(1-p)M(\rho_2)$, where $p \in(0,1).$\\
\indent It has been proved that (M2) and (M5) are equivalent to (M3)
and (M4)\cite{XSGJ}. We will call $M$ an
imaginarity monotone if it satisfies (M1), (M2) and (M4).\\
\indent Recently, many imaginarity measures have been proposed. The
relative entropy of imaginarity is defined by\cite{XSGJ}
\begin{equation}\label {eq3}
M_{r}(\rho)=S\bigg[\frac{1}{2}(\rho+\rho^T)\bigg]-S(\rho),
\end{equation}
Tsallis relative entropy of imaginarity is defined by  \cite{XJ}
\begin{equation}\label {eq4}
M_{T, u}(\rho)=1-\mathrm{tr}\left[\rho^u(\rho^*)^{1-u}\right],
\end{equation} where $u \in (0,1)$ and $\alpha$-$z$-R\'enyi relative entropy of imaginarity is defined by  \cite{CXLQ}
\begin{equation}\label {eq5}
M_{\alpha,z}^R(\rho)=1-f_{\alpha,z}(\rho,\rho^*),
\end{equation} where $f_{\alpha,z}(\rho, \sigma)=$tr$\left(\sigma^{\frac{1-\alpha}{2z}} \rho^{\frac{\alpha}{z}} \sigma^{\frac{1-\alpha}{2z} }\right)^z$ and $0<\max\{\alpha,1-\alpha\} \le z <1$. We denote $M_{\alpha,z}^R(\rho)$ by $M_{\alpha}^R(\rho)$ when $\alpha=z$.\\
\indent In addition, there are also many other imaginarity measures, such as Tsallis relative operator entropy of imaginarity \cite{CXLQ}, geometric imaginarity measure \cite{WKDK} and so on. They all have elegant properties.\\
\indent The unified $(\alpha,\beta)$-relative entropy has been introduced in \cite{WJJW}.\\
\indent {\bf Definition 4} \cite{WJJW} \quad For any $\alpha \in [0,1]$ and any $\beta \in \mathbb{R}$, the unified $(\alpha,\beta)$-relative entropy is defined by
\begin{equation}\label {eq6}
D_{\alpha}^{\beta}(\rho||\sigma)=\begin{cases}
  &H_{\alpha}^{\beta}(\rho||\sigma)  \qquad  \text{ if } 0\le \alpha<1 ,\beta \neq 0,\\
  &H_{\alpha}(\rho||\sigma)  \qquad \text{ if } 0 \le \alpha <1,\beta=0,\\
  &H^{\alpha}(\rho||\sigma)  \qquad \text{ if }  0 \le \alpha <1,\beta=1,\\
  &_{\frac{1}{\alpha}}H(\rho||\sigma) \qquad  \text{ if } 0<\alpha<1,\beta=\frac{1}{\alpha}, \\
  &H(\rho||\sigma) \qquad \quad \text{if } \alpha=1,
\end{cases}
\end{equation}\\  where
\begin{equation}\label {eq7}
H_{\alpha}^{\beta}(\rho||\sigma)=\frac{1}{(\alpha-1)\beta}[( \mathrm{tr}( \rho^{\alpha} \sigma^{1-\alpha}))^{\beta}-1],
\end{equation}
\begin{equation}\label {eq8}
H_{\alpha}(\rho||\sigma)=\frac{1}{(\alpha-1)} \log_{2}(\mathrm{tr} (\rho^{\alpha} \sigma^{1-\alpha})),
\end{equation}
\begin{equation}\label {eq9}
H^{\alpha}(\rho||\sigma)=\frac{1}{(\alpha-1)}[\mathrm{tr} (\rho^{\alpha} \sigma^{1-\alpha})-1],
\end{equation}
\begin{equation}\label {eq10}
_{\alpha}H(\rho||\sigma)=\frac{1}{(\alpha-1)}[( \mathrm{tr}( \rho^{\frac{1}{\alpha}} \sigma^{1-\frac{1}{\alpha}}))^{\alpha}-1],
\end{equation}
\begin{equation}\label {eq11}
H(\rho||\sigma)=\mathrm{tr}(\rho \log_{2}\rho)-\mathrm{tr}(\rho \log_{2} \sigma).
\end{equation}
\indent It can be easily obtained that $H_{\alpha}^{\beta}(\rho||\sigma)$ reduces to the R\'enyi relative entropy, Tsallis relative entropy and relative entropy when $\beta \to 0$, $\beta=1$ and $\alpha \to 1$, respectively.\\
\indent In order to facilitate the subsequent proofs of the theorems, we present the following lemmas.\\
\indent {\bf Lemma 1}\cite{WJJW}\quad The unified $(\alpha,\beta)$-relative entropy satisfies the following properties for any quantum states $\rho,\sigma$ and any $\alpha \in (0,1)$ and $\beta \in (0,1]:$\\
\indent (i) $D_{\alpha}^{\beta}(\rho|| \sigma) \ge 0 $ and $D_{\alpha}^{\beta}(\rho|| \sigma)=0$ iff $\rho=\sigma$. \\
\indent (ii) $D_{\alpha}^{\beta}(\mathcal{E} (\rho)|| \mathcal{E} (\sigma)) \le D_{\alpha}^{\beta}(\rho|| \sigma)$ whenever $\mathcal{E}$ is a quantum operation.\\
\indent (iii) $D_{\alpha}^{\beta}(\sum_{j} \lambda_{j} \rho_{j} || \sum_{j} \lambda_{j} \sigma_{j}) \le \sum_{j} \lambda_{j} D_{\alpha}^{\beta}(\rho_{j} ||\sigma_{j})$.\\
\indent (iv) For any states $\rho$ and $\sigma$ on $H_{1} \otimes H_{2}$, $D_{\alpha}^{\beta}(\rho_{1} ||\sigma_{1}) \le D_{\alpha}^{\beta}(\rho ||\sigma)$, where $\rho_{1}$ and $\sigma_{1}$ are the partial traces of $\rho$ and $\sigma$ on $H_{1}$, respectively.\\
\indent (v) $D_{\alpha}^{\beta}(\rho ||\sigma)$ is increasing with respect to $\alpha\in (0,1)$ for fixed $0<\beta\le1$.\\
\indent (vi) $ D_{\alpha}^{\beta}(\rho ||\sigma)$ is decreasing with respect to $\beta\in(0,1]$ for fixed $0<\alpha<1$. \\
\indent {\bf Lemma 2}\cite{XJ}\quad For any real operation $\mathcal {E}$ and any state $\rho$, it holds that
\begin{equation}\label {eq12}
\mathcal {E}(\rho^*)=[\mathcal {E}(\rho)]^{*}.
\end{equation}

\indent The following lemma plays an important role in the proof of Eq. (\ref{eq24}) in Example 1, and the proof of it is given in Appendix A.\\
\indent {\bf Lemma 3}\quad Let $A>\sqrt{B^2+C^2}$, $B^2+C^2>0$, $\alpha \in (0,1)$ and $f(x,\theta)=A[x^{1-\alpha}+(1-x)^{1-\alpha}]+(B\sin\theta+C\cos\theta)[(1-x)^{1-\alpha}-x^{1-\alpha}],$ where $x \in [0,\frac{1}{2}]$, $\theta \in [0,2\pi]$. Then $f(x,\theta)$ attains its maximum value at $(x_{0},\theta_{0})=\left(\frac{1}{\left(\frac{A+\sqrt{B^2+C^2}}{A-\sqrt{B^2+C^2}}\right)^{\frac{1}{\alpha}}+1},\arcsin \frac{B}{\sqrt{B^2+C^2}}\right)$.\\

\indent Finally, Lemma 4 is crucial in proving item
(i) of Theorem 9, and the proof is given in Appendix B.\\
\indent {\bf Lemma 4}\quad Suppose that $\sigma_{\rho}$ and $\sigma_{\tau}$ are the quantum state that reaches the minimum in Eq. (\ref{eq20}) when the input states are $\rho$ and $\tau$, respectively. Then $p\sigma_{\rho}\oplus (1-p)\sigma_{\tau}$ is a quantum state achieving the minimum in Eq. (\ref{eq20}) when the input state is $p\rho\oplus (1-p)\tau$.\\

\noindent {\bf 3 Imaginarity monotones based on the unified $(\alpha,\beta)$-relative entropy}\\\hspace*{\fill}\\
\indent In this section, we mainly investigate two different imaginarity monotones induced by the unified $(\alpha,\beta)$-relative entropy, along with its properties.\\
\indent Now, we define the first imaginarity quantifier induced by unified $(\alpha,\beta)$-relative entropy as\\
\begin{equation}\label {eq13}
M_{\alpha,\beta}^{H}(\rho)=\frac{1}{(\alpha-1)\beta}[( \mathrm{tr} (\rho^{\alpha} (\rho^*)^{1-\alpha}))^{\beta}-1],
\end{equation}  where $\alpha \in (0,1),\beta \in (0,1].$\\
\indent {\bf Theorem 1}\quad$M_{\alpha,\beta}^{H}(\rho)$ defined by Eq. (\ref{eq13}) is an imaginarity monotone.\\
\indent \textit {Proof}. Due to Lemma 1(i), we know that
$M_{\alpha,\beta}^{H}(\rho)=D_{\alpha}^{\beta}(\rho||\rho^*) \ge 0$
and
\begin{equation}
M_{\alpha,\beta}^{H}(\rho)=0 \Leftrightarrow \rho=\rho^*  \Leftrightarrow \rho \in \mathcal{F}. \notag
\end{equation} Thus, $M_{\alpha,\beta}^{H}(\rho)$ satisfies (M1).\\
\indent From Lemma 1(ii), we know $D_{\alpha}^{\beta}(\rho||\sigma)$
is nonincreasing under any quantum operation when $\alpha \in (0,1)$
and $\beta \in(0,1].$ Let $\mathcal{E}$ be any real operation.
Combining Lemma 1(ii) and Lemma 2, we have
\begin{align}
M_{\alpha,\beta}^{H}(\mathcal {E}(\rho))
=&\frac{1}{(\alpha-1)\beta}[(\mathrm{tr}(\mathcal{E}(\rho)^{\alpha} ((\mathcal{E}(\rho))^{*})^{1-\alpha}))^{\beta}-1] \notag \\
=&\frac{1}{(\alpha-1)\beta}[(\mathrm{tr}(\mathcal{E}(\rho)^{\alpha} (\mathcal{E}(\rho^{*}))^{1-\alpha}))^{\beta}-1]  \notag \\
=&D_{\alpha}^{\beta}(\mathcal{E}(\rho)||
\mathcal{E}(\rho^*))\notag\\
\leq &
D_{\alpha}^{\beta}(\rho|| \rho^*) \notag \\
 =& \frac{1}{(\alpha-1)\beta}[(\mathrm{tr}(\rho^{\alpha} (\rho^{*})^{1-\alpha}))^{\beta}-1] \notag \\
=&M_{\alpha,\beta}^{H}(\rho), \notag
\end{align}
which implies that (M2) holds.

\indent Let $p_{n} \ge 0$ with $\sum_{n}p_{n}=1$ and $\rho_{n} \in
\mathcal D(\mathcal H).$ From Lemma 1(iii), we get
\begin{align}
M_{\alpha,\beta}^{H}\left(\sum_{n} p_{n} \rho_{n}\right)
=&D_{\alpha}^{\beta}\left(\sum_{n} p_{n} \rho_{n} ||\sum_{n} p_{n} \rho_{n}^{*}\right) \notag \\
\le&  \sum_{n} p_{n} D_{\alpha}^{\beta}(\rho_{n}||\rho_{n}^*)  \notag \\
=&  \sum_{n} p_{n} M_{\alpha,\beta}^{H}(\rho_{n}). \notag
\end{align}
Hence, (M4) is proved. Therefore, $M_{\alpha,\beta}^{H}(\rho)$ is an imaginarity monotone. This completes the proof. $\hfill\qedsymbol$ \\\hspace*{\fill}\\
\indent Furthermore, when $\rho=| \psi \rangle \langle \psi|$ is a pure state, we have
\begin{align}\label{eq14}
M_{\alpha,\beta}^{H}( \rho)=M_{\alpha,\beta}^{H}( | \psi \rangle \langle \psi |)=&\frac{1}{(\alpha-1) \beta}[(\mathrm{tr}(|\psi \rangle \langle \psi| \psi^{*} \rangle \langle \psi^{*}|))^{\beta}-1] \notag \\
=&\frac{1}{(\alpha-1) \beta}[|\langle \psi|\psi^{*} \rangle|^{2\beta}-1].
\end{align}
\indent In order to measure super-fidelity and
sub-fidelity, a quantum network-based experimental scheme has been
designed, in which $|\Psi\rangle_{12}$ is the program state as a
mean value of $\sigma_z$ of the measured controlling qubit (see Fig.
4 in\cite{MJAP}). By setting $|\Psi\rangle_{12}=|0\rangle|0\rangle$
one gets $\mathrm{tr}(\rho_1\rho_2)$\cite{MJAP}. Noting that
$\mathrm{tr}(\rho_1\rho_2)=|\langle\psi|\psi^{*}\rangle|^2$ when
$\rho_1=|\psi\rangle\langle\psi|$ and
$\rho_2=|\psi^{*}\rangle\langle\psi^{*}|$, it can be seen that the
quantity in Eq. (\ref{eq14}), i.e., $M_{\alpha,\beta}^{H}(\rho)$ in
the pure state case, can be measured
using the scheme proposed in \cite{MJAP}.\\
\indent From Eq. (\ref{eq13}), we can deduce that when $\beta=1$, $M_{\alpha,\beta}^{H}(\rho)=\frac{1}{1-\alpha}M_{T, u}(\rho)$. Excluding the influence of coefficients, $M_{\alpha,\beta}^H(\rho)$ reduces to the imaginarity measure in \cite{XJ}. In addition, $M_{\alpha,\beta}^{H}(\rho)$ has many elegant properties.\\
\indent {\bf Theorem 2}\quad For any quantum state $\rho$ on $\mathcal H$, and $\alpha \in (0,1)$, $\beta \in(0,1]$, we have $\frac{1}{1-\alpha}M_{1-\alpha,\beta}^{H}( \rho)=\frac{1}{\alpha}M_{\alpha,\beta}^{H}( \rho).$\\
\indent  \textit {Proof}. Note that
\begin{align}
M_{1-\alpha,\beta}^{H}( \rho)
=&\frac{1}{-\alpha\beta}[(\mathrm{tr}(\rho^{1-\alpha}(\rho^{*})^{\alpha}))^{\beta}-1] \notag \\
=&\frac{1-\alpha}{\alpha} \cdot \frac{1}{(\alpha-1)\beta}[(\mathrm{tr}(\rho^{1-\alpha}(\rho^{*})^{\alpha}))^{\beta}-1] \notag \\
=&\frac{1-\alpha}{\alpha}M_{\alpha,\beta}^{H}( \rho^{*})=\frac{1-\alpha}{\alpha}M_{\alpha,\beta}^{H}( \rho). \notag
\end{align} So Theorem 2 holds. This completes the proof. $\hfill\qedsymbol$ \\\hspace*{\fill}\\
\indent {\bf Theorem 3}\quad (Superadditivity under direct sum) For any quantum state $\rho_1 $, $\rho_2$ on $\mathcal H$, $p \in [0,1]$ and $\alpha \in (0,1)$, $\beta \in(0,1]$, we have $M_{\alpha,\beta}^{H}( p \rho_{1} \oplus (1-p)\rho_{2}) \ge pM_{\alpha,\beta}^{H}( \rho_{1})+(1-p)M_{\alpha,\beta}^{H}( \rho_{2})$.\\
\indent  \textit {Proof}. Direct calculation shows that
\begin{align}
M_{\alpha,\beta}^{H}( p \rho_{1} \oplus (1-p)\rho_{2})
=&\frac{1}{(\alpha-1)\beta}[(\mathrm{tr}((p \rho_{1} \oplus (1-p)\rho_{2})^{\alpha}(p \rho_{1}^{*} \oplus (1-p)\rho_{2}^{*})^{1-\alpha}))^{\beta}-1] \notag \\
=&\frac{1}{(\alpha-1)\beta}[(\mathrm{tr}(p\rho_{1}^{\alpha}(\rho_{1}^{*})^{1-\alpha}\oplus (1-p)\rho_{2}^{\alpha}(\rho_{2}^{*})^{1-\alpha}))^{\beta}-1] \notag \\
=&\frac{1}{(\alpha-1)\beta}[(p\mathrm{tr}(\rho_{1}^{\alpha}(\rho_{1}^{*})^{1-\alpha})+(1-p)\mathrm{tr}(\rho_{2}^{\alpha}(\rho_{2}^{*})^{1-\alpha}))^{\beta}-1]  \notag
\end{align} and
\begin{align}
&pM_{\alpha,\beta}^{H}( \rho_{1})+(1-p)M_{\alpha,\beta}^{H}( \rho_{2}) \notag \\
&=\frac{1}{(\alpha-1)\beta}[p(\mathrm{tr}(\rho_{1}^{\alpha}(\rho_{1}^*)^{1-\alpha}))^{\beta}-p]
+\frac{1}{(\alpha-1)\beta}[(1-p)(\mathrm{tr}(\rho_{2}^{\alpha}(\rho_{2}^*)^{1-\alpha}))^{\beta}-(1-p)] \notag \\
&=\frac{1}{(\alpha-1)\beta}[p(\mathrm{tr}(\rho_{1}^{\alpha}(\rho_{1}^*)^{1-\alpha}))^{\beta}+(1-p)(\mathrm{tr}(\rho_{2}^{\alpha}(\rho_{2}^*)^{1-\alpha}))^{\beta}-1]. \notag
\end{align}
\indent Due to the concavity of $f(x)=x^\beta$, we obtain
\begin{align}\label{eq15}
(p\mathrm{tr}(\rho_{1}^{\alpha}(\rho_{1}^{*})^{1-\alpha})+(1-p)\mathrm{tr} (\rho_{2}^{\alpha}(\rho_{2}^{*})^{1-\alpha}))^{\beta} \ge p(\mathrm{tr}(\rho_{1}^{\alpha}(\rho_{1}^*)^{1-\alpha}))^{\beta}+(1-p)(\mathrm{tr}(\rho_{2}^{\alpha}(\rho_{2}^*)^{1-\alpha}))^{\beta}.
\end{align} Therefore, $M_{\alpha,\beta}^{H}( p \rho_{1} \oplus (1-p)\rho_{2}) \ge pM_{\alpha,\beta}^{H}( \rho_{1})+(1-p)M_{\alpha,\beta}^{H}( \rho_{2})$. This completes the proof. $\hfill\qedsymbol$ \\\hspace*{\fill}\\
\indent {\bf Corollary 1}\quad For $\alpha \in (0,1)$ and $\beta \in(0,1]$, $M_{\alpha,\beta}^{H}(\rho)$ satisfies the additivity under direct sum if and only if $\beta = 1.$\\
\indent  \textit {Proof}. From the proof of Theorem 3, it follows that the equality in Eq. (\ref{eq15}) holds iff $\mathrm{tr}(\rho_{1}^{\alpha}(\rho_{1}^{*})^{1-\alpha})=\mathrm{tr} (\rho_{2}^{\alpha}(\rho_{2}^{*})^{1-\alpha})$ or $\beta=1$. However, $\mathrm{tr}(\rho_{1}^{\alpha}(\rho_{1}^{*})^{1-\alpha})=\mathrm{tr} (\rho_{2}^{\alpha}(\rho_{2}^{*})^{1-\alpha})$ can not be ture for any quantum states  $\rho_1 $ and $\rho_2$. Therefore, the equality in Eq. (\ref{eq15}), and thus the additivity for $M_{\alpha,\beta}^{H}(\rho)$ holds iff $\beta=1$. This completes the proof. $\hfill\qedsymbol$ \\\hspace*{\fill}\\
\indent {\bf Theorem 4}\quad For quantum state $\rho$ on $\mathcal{H}_{1} \otimes \mathcal{H}_{2}$, and $\alpha \in(0,1)$, $\beta \in (0,1]$, we have $M_{\alpha,\beta}^{H}(\rho_{1}) \le M_{\alpha,\beta}^{H}( \rho)$, where $\rho_{1}$ is the partial trace of $\rho$ on $\mathcal{H}_{1}$. \\
\indent  \textit {Proof}. Note that $\rho_1^*$ is the partial trace of $\rho^*$ on $H_1$ if $\rho_1$ is the partial trace of $\rho$ on $H_1$. Using Lemma 1(iv), we obtain $D_{\alpha}^{\beta}(\rho_{1} ||\rho_{1}^*) \le D_{\alpha}^{\beta}(\rho ||\rho^*)$, that is, $M_{\alpha,\beta}^{H}( \rho_{1}) \le M_{\alpha,\beta}^{H}( \rho)$. This completes the proof. $\hfill\qedsymbol$ \\\hspace*{\fill}\\
\indent {\bf Theorem 5}\quad (Subadditivity under tensor product)
For any quantum state $\rho $, $\tau$ on $\mathcal H_A$ and
$\mathcal H_B$, respectively, and $\alpha \in (0,1)$, $\beta
\in(0,1]$, we have
\begin{align}\label{eq16}
M_{\alpha,\beta}^{H}( \rho \otimes \tau) \le M_{\alpha,\beta}^{H}( \rho)+M_{\alpha,\beta}^{H}( \tau).
\end{align}
\indent  \textit {Proof}. Note that
\begin{align}
M_{\alpha,\beta}^{H}( \rho \otimes \tau)
=&\frac{1}{(\alpha-1)\beta}[(\mathrm{tr}((\rho \otimes \tau)^{\alpha}((\rho \otimes \tau)^{*})^{1-\alpha}))^{\beta}-1] \notag \\
=&\frac{1}{(\alpha-1)\beta}[(\mathrm{tr}((\rho \otimes \tau)^{\alpha}(\rho^{*} \otimes \tau^{*})^{1-\alpha}))^{\beta}-1] \notag \\
=&\frac{1}{(\alpha-1)\beta}[(\mathrm{tr}((\rho^{\alpha} \otimes \tau^{\alpha})((\rho^{*})^{1-\alpha} \otimes (\tau^{*})^{1-\alpha})))^{\beta}-1] \notag \\
=&\frac{1}{(\alpha-1)\beta}[(\mathrm{tr}(\rho^{\alpha} (\rho^{*})^{1-\alpha})  \mathrm{tr}(\tau^{\alpha} (\tau^{*})^{1-\alpha}))^{\beta}-1] \notag \\
=&\frac{1}{(\alpha-1)\beta}[(\mathrm{tr}(\rho^{\alpha}(\rho^{*})^{1-\alpha}))^{\beta}-1]  [(\mathrm{tr}(\tau^{\alpha}(\tau^{*})^{1-\alpha}))^{\beta}-1] \notag \\
&+\frac{1}{(\alpha-1)\beta}[(\mathrm{tr}(\rho^{\alpha}(\rho^{*})^{1-\alpha}))^{\beta}-1]+\frac{1}{(\alpha-1)\beta}[(\mathrm{tr}(\tau^{\alpha}(\tau^{*})^{1-\alpha}))^{\beta}-1] \notag \\
=&(\alpha-1)\beta M_{\alpha,\beta}^{H}(\rho) M_{\alpha,\beta}^{H}(\tau)+M_{\alpha,\beta}^{H}( \rho)+M_{\alpha,\beta}^{H}(\tau). \notag
\end{align} Since $ M_{\alpha,\beta}^{H}( \rho)$, $ M_{\alpha,\beta}^{H}( \tau)\ge0$, and $\alpha \in (0,1)$, $\beta \in(0,1]$, we obtain $M_{\alpha,\beta}^{H}( \rho \otimes \tau) \le M_{\alpha,\beta}^{H}( \rho)+M_{\alpha,\beta}^{H}( \tau)$. This completes the proof. $\hfill\qedsymbol$ \\\hspace*{\fill}\\
\indent \textbf{Remark 1}. (1) From the proof of Theorem 5, it can be seen that the equality in Eq. (\ref{eq16}) holds iff $M_{\alpha,\beta}^{H}( \rho)=0$ or $ M_{\alpha,\beta}^{H}( \tau)=0$, i.e., iff $\rho \in  \mathcal F$ or $\tau \in  \mathcal F$. In this case, Eq. (\ref{eq16}) reduces to $M_{\alpha,\beta}^{H}( \rho \otimes \tau)=M_{\alpha,\beta}^{H}( \tau)$ or $M_{\alpha,\beta}^{H}( \rho \otimes \tau)=M_{\alpha,\beta}^{H}( \rho)$, which implies ancillary independence of the quantifier to some extent.\\
\indent (2) It is worth pointing out that Eq. (\ref{eq16}) does not hold for any bipartite quantum state $\rho^{ab}$. To see this, consider the bipartite state $\rho^{ab}=(|00\rangle\langle00|-\mathrm{i}|00\rangle\langle11|+\mathrm{i}|11\rangle\langle00|+|11\rangle\langle11|)/2$. It follows that $\rho^a=\rho^b=\frac{I}{2}\in  \mathcal F$. Then $M_{\alpha,\beta}^{H}( \rho^a)=M_{\alpha,\beta}^{H}( \rho^b)=0$ and $M_{\alpha,\beta}^{H}( \rho^{ab}) > 0$, which implies that $M_{\alpha,\beta}^{H}(\rho^{ab})> M_{\alpha,\beta}^{H}(\rho^a)+M_{\alpha,\beta}^{H}(\rho^b)$.\\
\indent (3) Note that the subadditivity under tensor
product is an important property, and it holds for the Tsallis
relative entropy of imaginarity, which has not been mentioned in
\cite{XJ}. In comparison, the property in Theorem 5 has not been
discussed for the trace norm of imaginarity, the fidelity of
imaginarity, the weight of imaginarity, the robustness of
imaginarity, the $l_1$ norm of imaginarity and the geometric-like
imaginarity in previous literatures.

\indent {\bf Theorem 6}\quad For any $\alpha \in(0,1)$ and $\beta \in (0,1]$, the imaginarity monotone $M_{\alpha,\beta}^{H}( \rho)$ satisfies the following monotonicity: \\
\indent (1) $M_{\alpha_{1},\beta}^{H}(\rho) \le M_{\alpha_{2},\beta}^{H}(\rho)$ for fixed $\beta$ when $\alpha_1 \le \alpha_2$. \\
\indent (2) $M_{\alpha,\beta_{1}}^{H}(\rho) \ge M_{\alpha,\beta_{2}}^{H}(\rho)$ for fixed $\alpha$ when $\beta_{1} \le   \beta_{2}$.\\
\indent  \textit {Proof}. (1) From Lemma 1(v), we have $ D_{\alpha_{1}}^{\beta}(\rho ||\sigma) \le D_{\alpha_{2}}^{\beta}(\rho ||\sigma)$ when $\beta > 0$ and $\alpha_{1} \le \alpha_{2}$. Letting $\sigma = \rho^{*}$, we have $M_{\alpha_{1},\beta}^{H}(\rho) \le M_{\alpha_{2},\beta}^{H}(\rho).$\\
\indent (2) From Lemma 1(vi), we have $ D_{\alpha}^{\beta_{1}}(\rho ||\sigma) \ge D_{\alpha}^{\beta_{2}}(\rho ||\sigma)$ when $\beta_{1} \le \beta_{2}$. Letting $\sigma = \rho^{*}$, we have $M_{\alpha,\beta_{1}}^{H}(\rho) \ge M_{\alpha,\beta_{2}}^{H}(\rho).$  This completes the proof. $\hfill\qedsymbol$ \\\hspace*{\fill}\\
\indent {\bf Theorem 7}\quad For any quantum state $\rho$, the order of $M_{\alpha,z}^{R}$, $M_{\alpha}^{R}$, $M_{\alpha}^{T}$ and $M_{\alpha,\beta}^{H}$ is as follows:
\begin{equation}\label {eq17}
M_{\alpha}^{R}(\rho) \le M_{\alpha,z}^{R}(\rho) \le M_{\alpha}^{T}(\rho) \le M_{\alpha,\beta}^{H}(\rho).
\end{equation}
\indent  \textit {Proof}. From \cite{CXLQ}, we know that
\begin{equation}\label {eq18}
M_{\alpha}^{R}(\rho) \le M_{\alpha,z}^{R}(\rho) \le M_{\alpha}^{T}(\rho).
\end{equation} Using Theorem 6(2), we have
\begin{equation}\label {eq19}
 M_{\alpha}^{T}(\rho) =M_{\alpha,1}^{H}(\rho) \le M_{\alpha,\beta}^{H}(\rho).
\end{equation} From Eqs. (\ref{eq18}-\ref{eq19}), we obtain Eq. (\ref{eq17}). This completes the proof. $\hfill\qedsymbol$ \\\hspace*{\fill}\\
\indent In quantum coherence theory
\cite{BTMC,YXDZ,BKSA,ML}, a common approach for constructing a
coherence quantifier for a quantum state $\rho$ is to minimize a
distance measure $D(\rho,\sigma)$ over all incoherent states
$\sigma$. A similar scheme can be used in quantum imaginary resource
theory. Now, we define the second imaginarity monotone induced by
unified $(\alpha,\beta)$-relative entropy as
\begin{equation}\label{eq20}
M_{\alpha,\beta}^{E}(\rho)=\min_{\sigma \in \mathcal F} \frac{1}{(\alpha-1)\beta}[(\mathrm{tr}(\rho^{\alpha} \sigma^{1-\alpha}))^{\beta}-1],
\end{equation} where $\alpha \in (0,1)$ and $\beta \in(0,1]$.\\
\indent {\bf Theorem 8}\quad $M_{\alpha,\beta}^{E}(\rho) $ defined by Eq. (\ref{eq20})
 is an imaginarity monotone.\\
\indent  \textit {Proof}. From Lemma 1(i), we can easily obtain that
$M_{\alpha,\beta}^{E}(\rho)$ satisfies (M1). For
(M2), it follows from Lemma 1(ii) that
\begin{align}
D_{\alpha}^{\beta}(\mathcal{E}(\rho)|| \mathcal{E}(\sigma)) \leq D_{\alpha}^{\beta}(\rho|| \sigma), \notag
\end{align}
holds for any real operation $\mathcal{E}$, which yields that
\begin{align}
\min_{\sigma \in \mathcal F}D_{\alpha}^{\beta}(\mathcal{E}(\rho)||
\mathcal{E}(\sigma)) \leq \min_{\sigma \in \mathcal F}
D_{\alpha}^{\beta}(\rho|| \sigma), \notag
\end{align}
Therefore, we have
\begin{align}
 M_{\alpha,\beta}^{E}(\mathcal {E}(\rho))
=& \min_{\sigma \in \mathcal F} \frac{1}{(\alpha-1)\beta}[(\mathrm {tr}(\mathcal{E}(\rho)^{\alpha} \sigma^{1-\alpha}))^{\beta}-1]  \notag \\
\le & \min_{\sigma \in \mathcal F} \frac{1}{(\alpha-1)\beta}[(\mathrm {tr}(\mathcal{E}(\rho)^{\alpha} (\mathcal{E}(\sigma))^{1-\alpha}))^{\beta}-1]  \notag \\
=& \min_{\sigma \in \mathcal F}
D_{\alpha}^{\beta}(\mathcal{E}(\rho)|| \mathcal{E}(\sigma)) \notag \\
\leq & \min_{\sigma \in \mathcal F} D_{\alpha}^{\beta}(\rho|| \sigma)  \notag \\
 =& \min_{\sigma \in \mathcal F} \frac{1}{(\alpha-1)\beta}[(\mathrm{tr}(\rho^{\alpha} \sigma^{1-\alpha}))^{\beta}-1]  \notag \\
=& M_{\alpha,\beta}^{E}(\rho).  \notag
\end{align}
\indent Let $p_{n} \ge 0$, $\sum_{n}p_{n}=1$ and $\rho_{n} \in
\mathcal D(\mathcal H).$ Then we have
\begin{align}
 M_{\alpha,\beta}^{E}\left(\sum_{n} p_{n} \rho_{n}\right)
=& \min_{\sigma \in \mathcal F}D_{\alpha}^{\beta}\left(\sum_{n} p_{n} \rho_{n}|| \sigma\right)  \notag \\
\le & D_{\alpha}^{\beta}\left(\sum_{n} p_{n} \rho_{n}||\sum_{n} p_{n} \sigma_{n}\right)  \notag \\
\le& \sum_{n}p_{n}D_{\alpha}^{\beta}(\rho_{n}||\sigma_{n})   \notag \\
=& \sum_{n} p_{n} M_{\alpha,\beta}^{E}(\rho_{n}), \notag
\end{align} where $M_{\alpha,\beta}^{E}(\rho_n)=D_{\alpha}^{\beta}(\rho_n||\sigma_n)$. This implies (M4) holds for $M_{\alpha,\beta}^{E}(\rho)$. This completes the proof. $\hfill\qedsymbol$ \\\hspace*{\fill}\\
\indent From Theorem 8, we can derive that when $\alpha \to 1 $, $M_{\alpha,\beta}^{E}(\rho)=M_{r}(\rho)$ and Theorem 2 degrades to the imaginarity measure induced by relative entropy in \cite{XSGJ}. In addition, $M_{\alpha,\beta}^{E}(\rho)$ also has many properties similar to $M_{\alpha,\beta}^{H}(\rho)$. \\
\indent {\bf Theorem 9}\quad For any $\alpha \in (0,1)$ and $\beta \in(0,1]$, $M_{\alpha,\beta}^{E}(\rho)$ satisfies the following properties:\\
\indent (1) (Superadditivity under direct sum) For any quantum state $\rho_1 $, $\rho_2$ on $\mathcal H$ and $p \in [0,1]$ , we have $M_{\alpha,\beta}^{E}( p \rho_{1} \oplus (1-p)\rho_{2}) \ge pM_{\alpha,\beta}^{E}( \rho_{1})+(1-p)M_{\alpha,\beta}^{E}( \rho_{2})$.\\
\indent (2) $M_{\alpha,\beta}^{E}(\rho)$ satisfies the additivity under direct sum if and only if $\beta = 1.$\\
\indent (3) For quantum state $\rho$ on $\mathcal{H}_{1} \otimes \mathcal{H}_{2}$, we have $M_{\alpha,\beta}^{E}(\rho_{1}) \le M_{\alpha,\beta}^{E}( \rho)$, where $\rho_{1}$ is the partial trace of $\rho$ on $\mathcal{H}_{1}$.\\
\indent (4) (Subadditivity under tensor product) For any quantum state $\rho$, $\tau$ on $\mathcal H_A$ and $\mathcal H_B$, respectively, we have $M_{\alpha,\beta}^{E}( \rho \otimes \tau) \le M_{\alpha,\beta}^{E}( \rho)+M_{\alpha,\beta}^{E}( \tau)$. \\
\indent (5) $M_{\alpha,\beta_{1}}^{E}(\rho) \ge M_{\alpha,\beta_{2}}^{E}(\rho)$ for fixed $\alpha$ when $\beta_{1} \le   \beta_{2}$.\\
\indent  \textit {Proof}. (1) Suppose that $\sigma_{1}$ and $\sigma_{2}$ are the quantum state that reaches the minimum in Eq. (\ref{eq20}) when the input states are $\rho_1$ and $\rho_2$, respectively. Utilizing Lemma 4, we obtain
\begin{align}
&pM_{\alpha,\beta}^{E}( \rho_{1})+(1-p)M_{\alpha,\beta}^{E}( \rho_{2}) \notag \\
&=\frac{1}{(\alpha-1)\beta}[p(\mathrm{tr}(\rho_{1}^{\alpha}(\sigma_{1})^{1-\alpha}))^{\beta}
+(1-p)(\mathrm{tr}(\rho_{2}^{\alpha}(\sigma_{2})^{1-\alpha}))^{\beta}-1] \notag \\
&\le\frac{1}{(\alpha-1)\beta}[(p\mathrm{tr}(\rho_{1}^{\alpha}(\sigma_{1})^{1-\alpha})+(1-p)\mathrm{tr}(\rho_{2}^{\alpha}(\sigma_{2})^{1-\alpha}))^{\beta}-1] \notag \\
&=\frac{1}{(\alpha-1)\beta}[(\mathrm{tr}((p \rho_{1} \oplus (1-p)\rho_{2})^{\alpha}(p \sigma_{1} \oplus (1-p)\sigma_{2})^{1-\alpha}))^{\beta}-1] \notag \\
&=M_{\alpha,\beta}^{E}(p\rho_1\oplus (1-p)\rho_2), \notag
\end{align} where the inequality is true because $f(x)=x^\beta$ is a convex function. So item (1) holds.\\
\indent (2) Imitating the proof of Corollary 1, item (2) can be easily obtained, and so we omit it here.\\
\indent (3) Let $\tilde{\sigma}$ be a quantum state achieving the
minimum in Eq. (\ref{eq20}), $\tilde{\sigma}_1$ be the partial trace
of $\tilde{\sigma}$ on $\mathcal{H}_1$ and $\sigma_1$ be any real
state on $\mathcal{H}_1$. Using Lemma 1(iv), we obtain
\begin{align}
M_{\alpha,\beta}^{E}(\rho)&=\min_{\sigma\in \mathcal{F}}\frac{1}{(\alpha-1)\beta}[(\mathrm{tr}(\rho^{\alpha} \sigma^{1-\alpha}))^{\beta}-1] \notag \\
&=\frac{1}{(\alpha-1)\beta}[(\mathrm{tr}(\rho^{\alpha} \tilde{\sigma}^{1-\alpha}))^{\beta}-1] \notag \\
&=D(\rho||\tilde{\sigma}) \notag \\
&\ge D(\rho_1||\tilde{\sigma}_1) \notag \\
&=\frac{1}{(\alpha-1)\beta}[(\mathrm{tr}(\rho_1^{\alpha} \tilde{\sigma}_1^{1-\alpha}))^{\beta}-1] \notag \\
&\ge \min_{\sigma_1 \in \mathcal{F}}\frac{1}{(\alpha-1)\beta}[(\mathrm{tr}(\rho_1^{\alpha} \sigma_1^{1-\alpha}))^{\beta}-1] \notag \\
&=M_{\alpha,\beta}^{E}(\rho_1), \notag
\end{align} that is, $M_{\alpha,\beta}^{E}(\rho_{1}) \le M_{\alpha,\beta}^{E}( \rho)$. Hence, item (3) is true.\\
\indent (4) Suppose that $\sigma_{\rho}$ and $\sigma_{\tau}$ are the
quantum state that reaches the minimum value in Eq. (\ref{eq20})
when the input states are $\rho$ and $\tau$, respectively. Then we
obtain
\begin{align}
M_{\alpha,\beta}^{E}( \rho \otimes \tau)
=&\min_{\sigma\in \mathcal{F}}\frac{1}{(\alpha-1)\beta}[(\mathrm{tr}((\rho \otimes \tau)^{\alpha}\sigma^{1-\alpha}))^{\beta}-1] \notag \\
\le&\min_{\sigma_a,\sigma_b\in \mathcal{F}}\frac{1}{(\alpha-1)\beta}[(\mathrm{tr}((\rho \otimes \tau)^{\alpha}(\sigma_a\otimes \sigma_b)^{1-\alpha}))^{\beta}-1] \notag \\
\le&\frac{1}{(\alpha-1)\beta}[(\mathrm{tr}((\rho \otimes \tau)^{\alpha}(\sigma_{\rho} \otimes \sigma_{\tau})^{1-\alpha}))^{\beta}-1] \notag \\
=&\frac{1}{(\alpha-1)\beta}[(\mathrm{tr}((\rho^{\alpha} \otimes \tau^{\alpha})(\sigma_{\rho}^{1-\alpha} \otimes \sigma_{\tau}^{1-\alpha})))^{\beta}-1] \notag \\
=&\frac{1}{(\alpha-1)\beta}[(\mathrm{tr}(\rho^{\alpha} \sigma_{\rho}^{1-\alpha})  \mathrm{tr}(\tau^{\alpha} \sigma_{\tau}^{1-\alpha}))^{\beta}-1] \notag \\
=&\frac{1}{(\alpha-1)\beta}[(\mathrm{tr}(\rho^{\alpha}\sigma_{\rho}^{1-\alpha}))^{\beta}-1]  [(\mathrm{tr}(\tau^{\alpha}\sigma_{\tau}^{1-\alpha}))^{\beta}-1] \notag \\
&+\frac{1}{(\alpha-1)\beta}[(\mathrm{tr}(\rho^{\alpha}\sigma_{\rho}^{1-\alpha}))^{\beta}-1]+\frac{1}{(\alpha-1)\beta}[(\mathrm{tr}(\tau^{\alpha}\sigma_{\tau}^{1-\alpha}))^{\beta}-1] \notag \\
=&(\alpha-1)\beta M_{\alpha,\beta}^{E}(\rho)
M_{\alpha,\beta}^{E}(\tau)+M_{\alpha,\beta}^{E}(
\rho)+M_{\alpha,\beta}^{E}(\tau). \notag
\end{align} Since $M_{\alpha,\beta}^{E}(\rho)$, $M_{\alpha,\beta}^{E}(\tau)\ge0$, and $\alpha\in(0,1)$, $\beta \in(0,1]$, we obtain $M_{\alpha,\beta}^{E}( \rho \otimes \tau) \le M_{\alpha,\beta}^{E}( \rho)+M_{\alpha,\beta}^{E}( \tau)$, which implies that item (4) holds.\\
\indent (5) Suppose that $\hat{\sigma}$ is a quantum state reaching
the minimum value in the definition of
$M_{\alpha,\beta_1}^{E}(\rho)$, that is,
\begin{equation}
M_{\alpha,\beta_1}^E(\rho)=\frac{1}{(\alpha-1)\beta_1}[(\mathrm{tr}(\rho^{\alpha} \hat{\sigma}^{1-\alpha}))^{\beta_1}-1]. \notag
\end{equation} Then it is obvious that
\begin{equation}
M_{\alpha,\beta_2}^E(\rho)=\frac{1}{(\alpha-1)\beta_2}[(\mathrm{tr}(\rho^{\alpha} \hat{\sigma}^{1-\alpha}))^{\beta_2}-1]. \notag
\end{equation} Using Lemma 1(vi), we have
\begin{equation}
M_{\alpha,\beta_1}^E(\rho)=D_{\alpha}^{\beta_1}(\rho\|\hat{\sigma})\ge
D_{\alpha}^{\beta_2}(\rho\|\hat{\sigma})=M_{\alpha,\beta_2}^E(\rho).
\notag
\end{equation} Therefore, item (5) is derived. This completes the proof.
$\hfill\qedsymbol$ \\\hspace*{\fill}\\
\indent \textbf{Remark 2}. It bears noting that $M_{\alpha_{1},\beta}^{E}(\rho) \le M_{\alpha_{2},\beta}^{E}(\rho)$ for fixed $\beta$ when $\alpha_1 \le \alpha_2$ is not true, since when $\alpha$ varies, the quantum state $\sigma$ that minimizes the quantity in Eq. (\ref{eq20}) may also change accordingly, making it impossible to compare the values of $M_{\alpha_{1},\beta}^{E}(\rho)$ and $M_{\alpha_{2},\beta}^{E}(\rho)$.\\

\noindent {\bf 4 Examples}\\\hspace*{\fill}\\
\indent In this section, we investigate the imaginarity monotones $M_{\alpha,\beta}^{H}(\rho)$ and $M_{\alpha,\beta}^{E}(\rho)$ under some special quantum states.\\
\indent  {\bf Example 1}\quad Consider a qubit state $\rho$
expressed in Bloch representation as
\begin{equation}\label {eq21}
\rho=\frac{1}{2}(\textit{I} + \vec r \cdot \vec \sigma)=\frac{1}{2}
\begin{pmatrix}
1+r_{3} & r_{1}-\mathrm{i}r_2\\
r_1+\mathrm{i}r_2 & 1-r_3\\
\end{pmatrix},
\end{equation} where $\vec r =(r_1,r_2,r_3)$ is a real vector with $r=|\vec r|=\sqrt{r_1^2+r_2^2+r_3^2} \le 1$ and $\vec \sigma=(\sigma_{x},\sigma_{y},\sigma_{z})$ with $\sigma_{x}=
\begin{pmatrix}
0 & 1\\
1 & 0\\
\end{pmatrix}$, $\sigma_{y}=
\begin{pmatrix}
0 & -\mathrm{i}\\
\mathrm{i} & 0\\
\end{pmatrix}$ and $\sigma_{z}=
\begin{pmatrix}
1 & 0\\
0 & -1\\
\end{pmatrix}$. If $\rho \in \mathcal{F}$, we obtain $M_{\alpha,\beta}^{H}(\rho)=0$. Otherwise, it follows from \cite{XJ}, we know that
\begin{align}\label{eq22}
\mathrm{tr}[\rho^{\alpha}(\rho^{*})^{1-\alpha}]
=&\frac{1}{2r^2}\Bigg \{  (1-r)\Bigg[\Big(r-\frac{r_2^2}{r-r_3}\Big)^{2} + \frac{r_1^{2} r_2^{2}}{(r-r_3)^2}\Bigg]+ (1+r)\Bigg[\Big(r- \frac{r_2^2}{r+r_3}\Big)^2 \notag \\
&+\frac{r_1^2 r_2^2}{(r+r_3)^2}\Bigg] + r_2^2[(1-r)^{\alpha}(1+r)^{1-\alpha}+(1-r)^{1-\alpha}(1+r)^{\alpha}]   \Bigg \}.
\end{align}
Therefore, by Eq. (\ref{eq13}), we have
\begin{align}\label{eq23}
M_{\alpha,\beta}^{H}(\rho)
=&\frac{1}{(\alpha-1)\beta}   \Bigg\{ \frac{1}{2^\beta r^{2\beta}}\Bigg\{  (1-r)\Bigg[\Big(r-\frac{r_2^2}{r-r_3}\Big)^{2} + \frac{r_1^{2} r_2^{2}}{(r-r_3)^2}\Bigg]+ (1+r)\Bigg[\Big(r- \frac{r_2^2}{r+r_3}\Big)^2 \notag \\
&+\frac{r_1^2 r_2^2}{(r+r_3)^2}\Bigg] +
r_2^2\Big[(1-r)^{\alpha}(1+r)^{1-\alpha}+(1-r)^{1-\alpha}(1+r)^{\alpha}\Big]
\Bigg\} ^{\beta}-1  \Bigg\}
\end{align}
for all $\alpha\in(0,1)$ and $\beta\in (0,1]$, and by Eq.
(\ref{eq20}), $M_{\alpha,\beta}^{E}(\rho)$ reads
\begin{align}\label {eq24}
M_{\alpha,\beta}^{E}(\rho)
=&\frac{1}{(\alpha-1)\beta} \left \{ \frac{1}{2^\beta} \left[ \left(1-\frac{1}{\left(\frac{(r-\sqrt{r_1^2+r_3^2})(1-r)^{\alpha}+(r+\sqrt{r_1^2+r_3^2})(1+r)^{\alpha}}{(r+\sqrt{r_1^2+r_3^2})(1-r)^{\alpha}+(r-\sqrt{r_1^2+r_3^2})(1+r)^{\alpha}}\right)^{\frac{1}{\alpha}}+1} \right)^{1-\alpha}    \notag \right.  \right. \\
&\left. \times
 \Bigg(\left(1-\frac{\sqrt{r_1^2+r_3^2}}{r}\right)\left(\frac{1-r}{2}\right)^{\alpha}+\left(1+\frac{\sqrt{r_1^2+r_3^2}}{r}\right)\left(\frac{1+r}{2}\right)^{\alpha}\Bigg)
\notag \right.\\
&\left.
+\left(\left(1+\frac{\sqrt{r_1^2+r_3^2}}{r}\right)  \left(\frac{1-r}{2}\right)^{\alpha}+\left(1-\frac{\sqrt{r_1^2+r_3^2}}{r}\right)
\left(\frac{1+r}{2}\right)^{\alpha}\right)\notag \right.\\
&\left. \left. \times \left(
\frac{1}{\left(\frac{(r-\sqrt{r_1^2+r_3^2})(1-r)^{\alpha}+(r+\sqrt{r_1^2+r_3^2})(1+r)^{\alpha}}{(r+\sqrt{r_1^2+r_3^2})(1-r)^{\alpha}+(r-\sqrt{r_1^2+r_3^2})(1+r)^{\alpha}}\right)^{\frac{1}{\alpha}}+1}
\right)^{1-\alpha} \right]^{\beta}-1\right \}
\end{align}
for all $\alpha\in(0,1)$ and $\beta\in (0,1]$.

\begin{figure}[H]\centering
\subfigure[]
{\begin{minipage}[figure1a]{0.49\linewidth}
\includegraphics[width=0.9\textwidth]{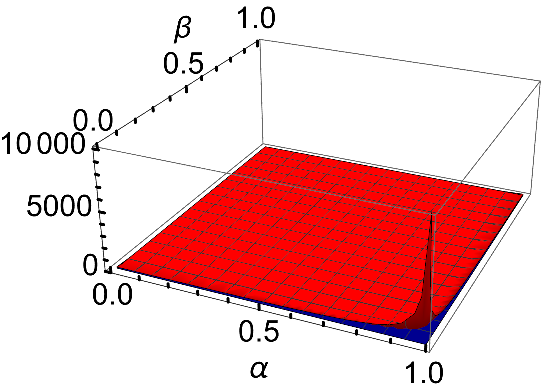}
\end{minipage}}
\subfigure[]
{\begin{minipage}[figure1b]{0.49\linewidth}
\includegraphics[width=0.9\textwidth]{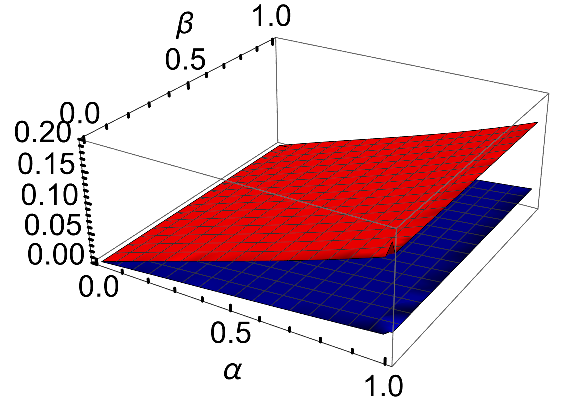}
\end{minipage}}
\caption{{The variations of $M^{H}_{\alpha,\beta}(\rho)$ and
$M^{E}_{\alpha,\beta}(\rho)$ in Eq. (\ref{eq23}) and Eq.
(\ref{eq24}) for fixed $\vec r$. Red (blue) surface represents the
value of $M^{H}_{\alpha,\beta}(\rho)$
($M^{E}_{\alpha,\beta}(\rho)$). $(\mathbf{a})$ $\vec r=(0,1,0)$
($\rho$ is a pure state); $(\mathbf{b})$ $\vec
r=(\frac{1}{2},\frac{1}{4},\frac{1}{2})$ ($\rho$ is a mixed state).
\label{fig:Fig1}}}
\end{figure}

\indent The proof of Eq. (\ref{eq24}) is given in Appendix C, and
the surfaces of $M_{\alpha,\beta}^{H}(\rho)$ and
$M_{\alpha,\beta}^{E}(\rho)$ in Eqs. (\ref{eq23}) and (\ref{eq24})
are plotted in Figure \ref{fig:Fig1}.
\indent \textbf{Remark 3}. From Appendix C, it can
be seen that Eq. (\ref{eq20}) achieves its minimum value when
\[
\sigma=\bar{\sigma}=\frac{1}{2}\begin{pmatrix}
  1 & 0 \\
 0 & 1
\end{pmatrix},
\] for any state
\[
\tilde{\rho}=\frac{1}{2}\begin{pmatrix}
  1 & -r_2 \mathrm{i} \\
  r_2 \mathrm{i} & 1
\end{pmatrix}
\] and
\begin{align}
\mathrm{tr}(\tilde{\rho}^\alpha
\bar{\sigma}^{1-\alpha})=\frac{1}{2}\left[(1-r_2)^{\alpha}+(1+r_2)^{\alpha}\right].
\notag
\end{align} By Eq. (\ref{eq22}), we have
\begin{align}
\mathrm{tr}(\tilde{\rho}^\alpha
(\tilde{\rho}^*)^{1-\alpha})=\frac{1}{2}\left[(1+r_2)^{1-\alpha}(1-r_2)^{\alpha}+(1+r_2)^{\alpha}(1-r_2)^{1-\alpha}\right].
\notag
\end{align} Let
\begin{align}
M(r_2,\alpha)&=\mathrm{tr}(\tilde{\rho}^\alpha (\tilde{\rho}^*)^{1-\alpha})-\mathrm{tr}(\tilde{\rho}^\alpha \bar{\sigma}^{1-\alpha}) \notag \\
&=\frac{1}{2}\left\{\left[(1+r_2)^{1-\alpha}-1\right](1-r_2)^{\alpha}+\left[(1-r_2)^{1-\alpha}-1\right](1+r_2)^{\alpha}\right\}.
\notag
\end{align} When $r_2 \neq\pm 1$, it can be shown that
\begin{align}
M_{r_2}^{'}(r_2,\alpha)=\frac{1}{2}\left\{(1-\alpha)\left[\frac{(1-r_2)^{\alpha}}{(1+r_2)^{\alpha}}-\frac{(1+r_2)^{\alpha}}{(1-r_2)^{\alpha}}\right]-\alpha\left[\frac{(1+r_2)^{1-\alpha}-1}{(1-r_2)^{1-\alpha}}-\frac{(1-r_2)^{1-\alpha}-1}{(1+r_2)^{1-\alpha}}\right]
\right \} \notag
\end{align} and
\begin{align}
M_{\alpha}^{'}(r_2,\alpha)=&\frac{1}{2}\Bigg\{(1+r_2)^{1-\alpha}(1-r_2)^{\alpha} \ln \frac{1-r_2}{1+r_2}+(1+r_2)^{\alpha}(1-r_2)^{1-\alpha}\ln \frac{1+r_2}{1-r_2} \notag \\
&-(1-r_2)^{\alpha} \ln (1-r_2)-(1+r_2)^{\alpha} \ln (1+r_2) \Bigg
\}. \notag
\end{align}
\indent Letting $M_{r_2}^{'}(r_2,\alpha)=0$ and
$M_{\alpha}^{'}(r_2,\alpha)=0$, we get $r_2=0$. Noting that
$M_{r_2}^{'}(r_2,\alpha)$ is symmetric about $r_2=0$,
$M_{r_2}^{'}(0,\alpha)=0$ and $M(\frac{1}{2},\alpha)=\frac{1}{2}(3^{\alpha}+3^{1-\alpha})-(\frac{1}{2})^{\alpha}-(\frac{3}{2})^{\alpha}\le M(0,\alpha)=0$ for any $\alpha \in (0,1)$, we obtain that $(0,\alpha)$ is a maximum value point. When $r_2=\pm 1$,
$M(-1,\alpha)=M(1,\alpha)=-2^{\alpha-1}<0$. Hence, $M(r_2,\alpha)\le 0$, that is, $M_{\alpha,\beta}^H(\tilde{\rho})\ge
M_{\alpha,\beta}^E(\tilde{\rho})$ for any $\alpha \in (0,1)$ and $\beta \in (0,1]$.\\
\indent We have not found a proof of $M_{\alpha,\beta}^H(\rho)\geq
M_{\alpha,\beta}^E(\rho)$ for any $\alpha\in (0,1)$, $\beta\in
(0,1]$ and any qubit state $\rho$, however, numerical experiments
shows that it might be true, and it seems difficult to find a
counterexample. It is also conjectured that
$M_{\alpha,\beta}^H(\rho)\geq M_{\alpha,\beta}^E(\rho)$ for any
$\alpha\in (0,1)$, $\beta\in (0,1]$ and any quantum state $\rho$.

\indent  {\bf Example 2}\quad Consider the modified Werner state
\begin{equation}\label {eq25}
\rho_\mathrm{w}=
\begin{pmatrix}
 \frac{1-k}{4} & 0 & 0 & 0 \\
 0 & \frac{1+k}{4} & \frac{k}{2}\mathrm{i} & 0 \\
 0 & -\frac{k}{2}\mathrm{i} & \frac{1+k}{4} & 0 \\
 0 & 0 & 0 & \frac{1-k}{4} \\
\end{pmatrix},
\end{equation}  where $k\in[0,1]$. Note that $\rho_\mathrm{w}$ is a pure state when $k=1$, $\rho_\mathrm{w} \in \mathcal F$ when
$k=0$ and the eigenvalues of $\rho_\mathrm{w}$ are $w_1=\frac{1-k}{4}$ and $w_2=\frac{1+3k}{4}$. By Eq. (\ref{eq13}), we have
\begin{align}\label {eq26}
M_{\alpha,\beta}^{H}(\rho_\mathrm{w})
=&\frac{1}{(\alpha-1)\beta}
\left\{\left[\frac{1}{32}(  w_1^{1-\alpha}w_2^{\alpha}+ w_1^{\alpha}w_2^{1-\alpha}-8(w_2-w_1)+24)\right]^{\beta}-1\right\}  \notag \\
=&\frac{1}{(\alpha-1)\beta}
\left\{\left[\frac{1}{8}( (3 k+1)^{\alpha } (1-k)^{1-\alpha }+(3
k+1)^{1-\alpha } (1-k)^{\alpha }-2 k+6)\right]^{\beta}-1\right\}
\end{align}
for all $\alpha\in(0,1)$ and $\beta\in (0,1]$.

\begin{figure}[H]\centering
{\begin{minipage}[figure2]{0.49\linewidth}
\includegraphics[width=1.0\textwidth]{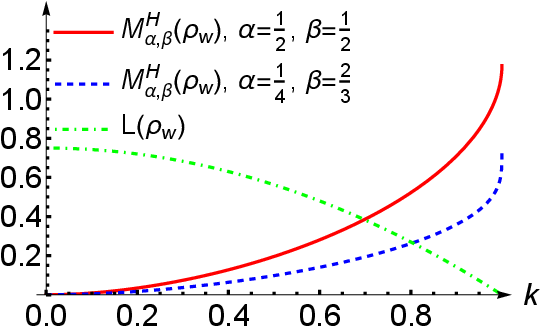}
\end{minipage}}
\caption{{The variations of $M_{\alpha,\beta}^{H}(\rho_\mathrm{w})$
and $L(\rho_w)$ with $k$ for fixed $\alpha$ and $\beta$. Red solid
line fixs $\alpha=\beta=\frac{1}{2}$ in Eq. (\ref{eq26}), while blue
dashed line fixs $\alpha=\frac{1}{4}$ and $\beta=\frac{2}{3}$ in Eq.
(\ref{eq26}), and green dot-dashed line represents $L(\rho_w)$.
\label{fig:Fig2}}}
\end{figure}

\indent And the linear entropy (mixedness) of the modified
Werner state $\rho_w$ is
$L(\rho_\mathrm{w})=1-\mathrm{tr}\rho_{\mathrm{w}}^2=3w_1(1+w_2-w_1)=\frac{3}{4}(1-k^2)$. We plot the variation of
$M_{\alpha,\beta}^{H}(\rho_\mathrm{w})$ and $L(\rho_\mathrm{w})$ with $k$ in Figure
\ref{fig:Fig2}. Interestingly,
$M_{\alpha,\beta}^{H}(\rho_\mathrm{w})$ is increasing with respect to $k$ while $L(\rho_\mathrm{w})$ is decreasing with respect to $k$. $M_{\alpha,\beta}^{H}(\rho_\mathrm{w})$ reaches its maximum value of
$\frac{1}{(\alpha-1)\beta}(\frac{1}{2^\beta}-1)$ for any fixed
$\alpha$ and $\beta$ when $k=1$, in which case $L(\rho_\mathrm{w})$
reaches its minimum value of $0$, while
$M_{\alpha,\beta}^{H}(\rho_\mathrm{w})$ reaches its minimum value of
$0$ when $k=0$ for any fixed $\alpha$ and $\beta$, in which case the
linear entropy (mixedness) $L(\rho_\mathrm{w})$ attains its maximum
value of $\frac{3}{4}$.
Then consider the modified isotropic state
\begin{equation}\label {eq27}
\rho_\mathrm{iso}=\left(
\begin{array}{cccc}
 \frac{1}{6}(2F+1) & 0 & 0 & \frac{1}{6}(4F-1)\mathrm{i} \\
 0 & \frac{1}{3}(1-F)  &  0  & 0 \\
 0 &0  & \frac{1}{3}(1-F)  & 0 \\
 -\frac{1}{6}(4F-1)\mathrm{i} & 0 & 0 & \frac{1}{6}(2F+1) \\
\end{array}
\right) ,
\end{equation}  where $F\in[0,1]$. Swapping the position of basis $|00\rangle\langle11|$ with $|01\rangle\langle10|$ and the position of basis $|11\rangle\langle 00|$ with $|10\rangle\langle01|$, and setting $F=\frac{3k+1}{4}$, we have $\rho_\mathrm{iso}=\rho_\mathrm{w}$, i.e, $\rho_\mathrm{iso}$ and $\rho_\mathrm{w}$ are essentially the same state, and we have $M_{\alpha,\beta}^{H}(\rho_\mathrm{iso})=M_{\alpha,\beta}^{H}(\rho_\mathrm{w})$ when $F=\frac{3k+1}{4}$.

\vskip0.1in

\noindent {\bf 5. Conclusions}\\\hspace*{\fill}\\
\indent We have proposed two classes of imaginarity monotones induced by the unified $(\alpha,\beta)$-relative entropy. The first class of imaginarity monotone $M_{\alpha,\beta}^{H}(\rho)$ is defined by setting $\sigma=\rho^*$. We have investigated several properties of $M_{\alpha,\beta}^{H}(\rho)$, including superadditivity under direct, subadditivity under tensor product and monotonicity, among others. The second class of imaginarity monotone $M_{\alpha,\beta}^{E}(\rho)$ is defined by minimizing the unified $(\alpha,\beta)$-relative entropy over all free states. The properties of  $M_{\alpha,\beta}^{E}(\rho)$ exhibit a high degree of similarity to those of $M_{\alpha,\beta}^{H}(\rho)$. Furthermore, we have presented the analytical expressions of $M_{\alpha,\beta}^{H}(\rho)$ for several special quantum states. For $M_{\alpha,\beta}^{E}(\rho)$, we have focused on deriving its analytical expression for qubit states, which differs significantly from the one of $M_{\alpha,\beta}^{H}(\rho)$ under the qubit state case.\\
\vskip0.1in

\noindent

\subsubsection*{Acknowledgements}
The authors would like to express their sincere gratitude to the anonymous referees for their useful suggestions which greatly improved this paper. The authors would also like to thank Prof. Jianwei Xu and Lin Zhang for fruitful discussions. This work was supported by National Natural Science Foundation of China (Grant No. 12161056) and Natural Science Foundation of Jiangxi Province (Grant No. 20232ACB211003).\\

\begin{center}
\textbf{Appendix A. Proof of Lemma 3}
\end{center}
\renewcommand{\theequation}{A\arabic{equation}}
\setcounter{equation}{0} 

\indent  We first prove that
$(x_{0},\theta_{0})=\left(\frac{1}{\left(\frac{A+\sqrt{B^2+C^2}}{A-\sqrt{B^2+C^2}}\right)^{\frac{1}{\alpha}}+1},\arcsin
\frac{B}{\sqrt{B^2+C^2}}\right)$ is the unique extreme point of
$f(x,\theta)$. In fact, it can be shown that
\begin{equation}
 f_{x}^{'}(x,\theta)=A(1-\alpha)\left(\frac{1}{x^{\alpha}}-\frac{1}{(1-x)^{\alpha}}\right)-(B\sin\theta +C\cos\theta)(1-\alpha)\left(\frac{1}{(1-x)^{\alpha}}+\frac{1}{x^{\alpha}}\right)
\notag
\end{equation}
and
\begin{equation}
f_{\theta}^{'}(x,\theta)=(B\cos\theta-C\sin\theta)[(1-x)^{1-\alpha}-x^{1-\alpha}] \notag .
\end{equation}
\indent Letting $f_{x}^{'}(x,\theta)=0$ and $f_{\theta}^{'}(x,\theta)=0$, we obtain that $x_0=\frac{1}{\left(\frac{A+\sqrt{B^2+C^2}}{A-\sqrt{B^2+C^2}}\right)^{\frac{1}{\alpha}}+1}$, $\sin\theta_0=\frac{B}{\sqrt{B^2+C^2}}$ and $\cos\theta_0=\frac{C}{\sqrt{B^2+C^2}}$, or $x_1=\frac{1}{\left(\frac{A-\sqrt{B^2+C^2}}{A+\sqrt{B^2+C^2}}\right)^{\frac{1}{\alpha}}+1}$, $\sin\theta_0=-\frac{B}{\sqrt{B^2+C^2}}$ and $\cos\theta_0=-\frac{C}{\sqrt{B^2+C^2}}$. Since $A-\sqrt{B^2+C^2} > 0$ and $x_0 \in [0,\frac{1}{2}]$ , $x_1 \notin [0,\frac{1}{2}]$. Therefore, $(x_{0},\theta_{0})$ is the unique extreme point of $f(x,\theta)$.\\
\indent Then we prove that $(x_{0},\theta_{0})$ is the maximum value
point of $f(x,\theta)$. Direct calculation shows that
\begin{equation}
f_{xx}^{''}(x,\theta)=-\alpha(1-\alpha)\left[(A-B\sin\theta-C\cos\theta)\frac{1}{x^{\alpha+1}}+(A+B\sin\theta+C\cos\theta)\frac{1}{(1-x)^{\alpha+1}}\right] \notag ,
\end{equation}
\begin{equation}
 f_{x\theta}^{''}(x,\theta)=-(1-\alpha)(B\cos\theta-C\sin\theta)\left[\frac{1}{(1-x)^{\alpha}}+\frac{1}{x^{\alpha}}\right] \notag
\end{equation}
and
\begin{equation}
f_{\theta\theta}^{''}(x,\theta)=-(B\sin\theta+C\cos\theta)[(1-x)^{1-\alpha}-x^{1-\alpha}] \notag .
\end{equation}
\indent It can be seen that $f_{xx}^{''}(x_{0},\theta_{0})<0$, $f_{x\theta}^{''}(x_{0},\theta_{0})=0$ and $f_{\theta\theta}^{''}(x_{0},\theta_{0})<0$. Therefore, $[f_{x\theta}^{''}(x_{0},\theta_{0})]^2-f_{xx}^{''}(x_{0},\theta_{0})f_{\theta\theta}^{''}(x_{0},\theta_{0})<0$. Noting that $f(x,\theta)$ is a continuous function, we conclude that $(x_{0},\theta_{0})$ is the maximum value point of $f(x,\theta)$. This completes the proof. $\hfill\qedsymbol$ \\\hspace*{\fill}\\

\begin{center}
\textbf{Appendix B. Proof of Lemma 4}
\end{center}
\indent \indent First of all, since $\sigma_{\rho}$
and $\sigma_{\tau}\in \mathcal{F}$, we have $p\sigma_{\rho}\oplus
(1-p)\sigma_\tau \in \mathcal{F}$. Suppose that
\begin{equation}
 M_{\alpha,\beta}^{E}(p\rho\oplus (1-p)\tau)=\frac{1}{(\alpha-1)\beta}[(\mathrm{tr}((p\rho\oplus (1-p)\tau)^{\alpha}\sigma_0^{1-\alpha}))^{\beta}-1]. \notag
\end{equation} Let
\[
\sigma_0^{1-\alpha}=\begin{pmatrix}
  \boldsymbol{S} &\boldsymbol{U}  \\
  \boldsymbol{V} &\boldsymbol{W}
\end{pmatrix}.
\]
Then we have
\begin{align}
\mathrm{tr}((p\rho\oplus (1-p)\tau)^{\alpha}\sigma_{0}^{1-\alpha})
&= \text{tr}\left[\begin{pmatrix}
                        (p\rho)^{\alpha} & \boldsymbol{0} \\
                       \boldsymbol{0} & ((1-p)\tau)^{\alpha}
                      \end{pmatrix}\begin{pmatrix}
                      \boldsymbol{S} &\boldsymbol{U}  \\
 \boldsymbol{V} &\boldsymbol{W} \notag
                      \end{pmatrix}\right]\\
               &= \mathrm{tr}(\boldsymbol{S}(p\rho)^{\alpha}+\boldsymbol{W}((1-p)\tau)^{\alpha}). \notag
\end{align}
Therefore, $\mathrm{tr}[(p\rho\oplus
(1-p)\tau)^{\alpha}\sigma_{0}^{1-\alpha}]$ attains its maximum value
when $\boldsymbol{S} =(p\sigma_\rho)^{1-\alpha}$ and
$\boldsymbol{W}=((1-p)\sigma_\tau)^{1-\alpha}$, which is independent of
$\boldsymbol{U}$ and $\boldsymbol{V}$, that is, when
\[
\sigma_0^{1-\alpha}=\begin{pmatrix}
  (p\sigma_\rho)^{1-\alpha} & \boldsymbol{0} \\
  \boldsymbol{0} & ((1-p)\sigma_\tau)^{1-\alpha}
\end{pmatrix}.
\] Thus, $\sigma_0=p\sigma_\rho\oplus (1-p)\sigma_\tau$ is a quantum state that reaches the minimum value in Eq.(\ref{eq20}) when the input state is $p\rho\oplus (1-p)\tau$. This completes the proof. $\hfill\qedsymbol$ \\\hspace*{\fill}\\
\begin{center}
\textbf{Appendix C. Proof of Eq. (\ref{eq24})}
\end{center}
\renewcommand{\theequation}{C\arabic{equation}}
\setcounter{equation}{0} 

\indent Let $\lambda_{1,2}=(1 \mp  r)/2$ be the eigenvalues of a
qubit state $\rho$, where $r=|\vec r|$. For any qubit free state
$\sigma$, we assume its Bloch vector is $\vec s=(s_{1},0,s_{3})$,
and the eigenvalues of $\sigma$ are $\mu_{1,2}=(1 \mp s)/2$, where
$s=|\vec s|$. If $r=0$, it is obvious that
$M_{\alpha,\beta}^E(\rho)=0$. Now we consider the case of $r\neq
0$. It follows from \cite{HAOY} that
\begin{equation}\label {eq:C1}
 \rho^{\alpha}=
\begin{pmatrix}
\frac{\lambda_{1}^{\alpha}+\lambda_{2}^{\alpha}}{2}+\frac{r_{3}(\lambda_{2}^{\alpha}-\lambda_{1}^{\alpha})}{2r} & \frac{(-r_{1}+\mathrm{i}r_{2})(\lambda_{1}^{\alpha}-\lambda_{2}^{\alpha})}{2r}\\
\frac{(-r_{1}-\mathrm{i}r_{2})(\lambda_{1}^{\alpha}-\lambda_{2}^{\alpha})}{2r} & \frac{\lambda_{1}^{\alpha}+\lambda_{2}^{\alpha}}{2}-\frac{r_{3}(\lambda_{2}^{\alpha}-\lambda_{1}^{\alpha})}{2r}\\
\end{pmatrix}
\end{equation} and
\begin{equation}\label {eq:C2}
 \sigma^{1-\alpha}=
\begin{pmatrix}
\frac{\mu_{1}^{1-\alpha}+\mu_{2}^{1-\alpha}}{2}+\frac{s_{3}(\mu_{2}^{1-\alpha}-\mu_{1}^{1-\alpha})}{2s} & \frac{-s_{1}(\mu_{1}^{1-\alpha}-\mu_{2}^{1-\alpha})}{2s}\\
\frac{-s_{1}(\mu_{1}^{1-\alpha}-\mu_{2}^{1-\alpha})}{2s} & \frac{\mu_{1}^{1-\alpha}+\mu_{2}^{1-\alpha}}{2}-\frac{s_{3}(\mu_{2}^{1-\alpha}-\mu_{1}^{1-\alpha})}{2s}\\
\end{pmatrix},
\end{equation} where $s\neq 0$. It implies that
\begin{align}\label{eq:C3}
\mathrm {tr}(\rho^{\alpha}\sigma^{1-\alpha})
=& \frac{1}{2}(\lambda_{1}^{\alpha}+\lambda_{2}^{\alpha})(\mu_{1}^{1-\alpha}+\mu_{2}^{1-\alpha})+\frac{r_{3}}{2r}(\lambda_{2}^{\alpha}-\lambda_{1}^{\alpha}) \cdot \frac{s_{3}}{s}(\mu_{2}^{1-\alpha}-\mu_{1}^{1-\alpha})  \notag \\
&+\frac{r_{1}}{2r}(\lambda_{2}^{\alpha}-\lambda_{1}^{\alpha}) \cdot \frac{s_{1}}{s}(\mu_{2}^{1-\alpha}-\mu_{1}^{1-\alpha}).
\end{align}
\indent Letting
$A=\frac{1}{2}(\lambda_{1}^{\alpha}+\lambda_{2}^{\alpha})$,
$B=\frac{r_{1}}{2r}(\lambda_{2}^{\alpha}-\lambda_{1}^{\alpha})$ and
$C=\frac{r_{3}}{2r}(\lambda_{2}^{\alpha}-\lambda_{1}^{\alpha})$,
$s_{1}=c \sin \theta$ and $s_{3}=c \cos \theta$, where $c \in [0,1]$
and $\theta \in [0,2\pi]$ and $x=\frac{1-c}{2}$, by Eq.
(\ref{eq:C3}) we have
\begin{align}\label {eq:C4}
\mathrm {tr}(\rho^{\alpha}\sigma^{1-\alpha})
=&A(x^{1-\alpha}+(1-x)^{1-\alpha})+(B\sin \theta+C\cos \theta)[(1-x)^{1-\alpha}-x^{1-\alpha}].
\end{align} And when $s=0$,
\begin{align}\label{eq:C5}
\mathrm
{tr}(\rho^{\alpha}\sigma^{1-\alpha})=\frac{\lambda_{1}^{\alpha}+\lambda_{2}^{\alpha}}{2^{1-\alpha}}.
\end{align}
\indent If $B^2+C^2=0$, we have $\mathrm {tr}(\rho^{\alpha}\sigma^{1-\alpha})_{\mathrm{max}}=\mathrm{max} \{A(x^{1-\alpha}+(1-x)^{1-\alpha}),(\lambda_{1}^{\alpha}+\lambda_{2}^{\alpha})/2^{1-\alpha}\}$. Direct calculation shows that $\mathrm {tr}(\rho^{\alpha}\sigma^{1-\alpha})$ achieves its maximum value of $2^{\alpha}A$ when $s_1=s_3=0$. In this case, $M_{\alpha,\beta}^E(\rho)=\frac{1}{(\alpha-1)\beta}(2^{\alpha \beta}A^{\beta}-1)$.\\
\indent If $B^2+C^2>0$, we have
$(\frac{1}{2})^{1-\alpha}(\lambda_{1}^{\alpha}+\lambda_{2}^{\alpha})\le
\mathrm{max}\{A(x^{1-\alpha}+(1-x)^{1-\alpha})+(B\sin \theta+C\cos
\theta)[(1-x)^{1-\alpha}-x^{1-\alpha}]\}$. So $\mathrm
{tr}(\rho^{\alpha}\sigma^{1-\alpha})$ achieves its maximum value
when $s\neq 0$. Note that $x \in [0,\frac{1}{2}]$,
$A>\sqrt{B^2+C^2}$, $B^2+C^2>0$ and $\alpha \in (0,1)$. It follows
from Lemma 3 that $\mathrm {tr}(\rho^{\alpha}\sigma^{1-\alpha})$
reaches its maximum value at
$(x_0,\theta_0)=\left(\frac{1}{\left(\frac{A+\sqrt{B^2+C^2}}{A-\sqrt{B^2+C^2}}\right)^{\frac{1}{\alpha}}+1},\arcsin
\frac{B}{\sqrt{B^2+C^2}}\right)$ and the maximum value is
\begin{align}\label {eq:C6}
\mathrm {tr}(\rho^{\alpha}\sigma^{1-\alpha})_{\max}
=(A+\sqrt{B^2+C^2})\left(1-\frac{1}{\left(\frac{A+\sqrt{B^2+C^2}}{A-\sqrt{B^2+C^2}}\right)^{\frac{1}{\alpha}}+1}\right)^{1-\alpha} \notag \\
+(A-\sqrt{B^2+C^2})\left( \frac{1}{\left(\frac{A+\sqrt{B^2+C^2}}{A-\sqrt{B^2+C^2}}\right)^{\frac{1}{\alpha}}+1} \right)^{1-\alpha},
\end{align} when $s\neq 0$. \\
\indent Since $\alpha \in (0,1)$ and $\beta \in(0,1]$, it is obvious that $M_{\alpha,\beta}^{E}(\rho)$ reaches its minimum value when $\mathrm {tr} (\rho^\alpha \sigma^{1-\alpha})$ reaches its maximum value. By Eq. (\ref{eq:C6}) and Eq. (\ref{eq20}), we obtain Eq. (\ref{eq24}). When $x_0=\frac{1}{\left(\frac{A+\sqrt{B^2+C^2}}{A-\sqrt{B^2+C^2}}\right)^{\frac{1}{\alpha}}+1}$, we have $c_0=1-2x_0=1-\frac{2}{\left(\frac{A+\sqrt{B^2+C^2}}{A-\sqrt{B^2+C^2}}\right)^{\frac{1}{\alpha}}+1}$, which implies that $\mathrm {tr}(\rho^{\alpha}\sigma^{1-\alpha})$ reaches its maximum value, and thus Eq. (\ref{eq20}) reaches its minimum value for qubit free state when $s_{1}=c_0 \sin \theta_0=\left(1-\frac{2}{\left(\frac{A+\sqrt{B^2+C^2}}{A-\sqrt{B^2+C^2}}\right)^{\frac{1}{\alpha}}+1}\right)\frac{B}{\sqrt{B^{2}+C^{2}}}$ and $s_{3}=c_0\cos \theta_0=-\left(1-\frac{2}{\left(\frac{A+\sqrt{B^2+C^2}}{A-\sqrt{B^2+C^2}}\right)^{\frac{1}{\alpha}}+1}\right)\frac{C}{\sqrt{B^{2}+C^{2}}}$. Note that the value of $\mathrm {tr}(\rho^{\alpha}\sigma^{1-\alpha})_{\mathrm{max}}$ when $B^2+C^2=0$ is equal to the value of $\mathrm {tr}(\rho^{\alpha}\sigma^{1-\alpha})_{\mathrm{max}}$ when $B^2+C^2\neq 0$. This completes the proof. $\hfill\qedsymbol$ \\\hspace*{\fill}\\

\small {}

\end{document}